\documentclass[twocolumn,aps,prd,amsmath,amssymb,preprintnumbers,longbibliography]{revtex4-1}
\usepackage{amsmath} \usepackage{amsfonts} \usepackage{amssymb}
\usepackage{bbm}
\usepackage{epsfig}
\usepackage{graphics}
\usepackage{graphicx}
\usepackage{titlesec}
\usepackage{mathtools}
\usepackage{environ}
\usepackage{dsfont}
\usepackage{version}
\usepackage[colorlinks=true]{hyperref}
\hypersetup{urlcolor=black,linkcolor=black,citecolor=black}
\usepackage{natbib}
\usepackage[toc,page]{appendix}

\usepackage{enumerate}

\usepackage{dsfont}

\makeatletter
\renewcommand*\env@matrix[1][c]{\hskip -\arraycolsep
  \let\@ifnextchar\new@ifnextchar
  \array{*\c@MaxMatrixCols #1}}
\makeatother

\makeatletter										
\def\p@subsection{\thesection .\,} 
\makeatother                                       

\newcommand{\be}{\begin{equation}}
\newcommand{\ee}{\end{equation}}
\newcommand{\ba}{\begin{align}}
\newcommand{\ea}{\end{align}}
\newcommand{\nn}{\nonumber}

\newcommand{\gl}{\big(}
\newcommand{\gr}{\big)}

\titleformat{\subsection}[block]{\normalfont\bfseries}{\thesubsection.}{1ex}{}
\titlespacing{\subsection}{0pt}{10pt}{1pt}[0pt]
\titleformat*{\section}{\large\bfseries}
\renewcommand{\thesubsection}{\arabic{subsection}}

\pdfoutput=1 

\usepackage{todonotes}
\usepackage{braket}

\newcommand{\tr}{\mathrm{tr}}

\newcommand{\mn}{\mu\nu}

\newcommand{\bel}[1]{\be\label{#1}}

\newcommand{\gbar}{\bar{g}}

\newcommand{\subt}[1]{_{\text{#1}}}

\newcommand{\qq}[1]{``#1"}

\newcommand{\eps}{\varepsilon}
\newcommand{\herm}{^{\dagger}}

\newcommand{\qtil}{\tilde{q}}

\newcommand{\2}{^{2}}
\newcommand{\kk}{_{k}}
\newcommand{\mo}{^{\mu}}
\newcommand{\no}{^{\nu}}
\newcommand{\mb}{_{\mu}}
\newcommand{\nb}{_{\nu}}
\newcommand{\mno}{^{\mu\nu}}
\newcommand{\mnb}{_{\mu\nu}}

\newcommand{\lambdh}{\lambda\subt{H}}
\newcommand{\Mbar}{\overline{M}}
\newcommand{\rhotil}{\tilde\rho}
\newcommand{\Phat}{\widehat{P}}

\definecolor{refkey}{rgb}{0,0,1}
\definecolor{labelkey}{rgb}{0,1,0}

\begin{document}


\title{\LARGE Scaling solution for field-dependent gauge couplings in quantum gravity}

\author{C. Wetterich}

\affiliation{Institut  f\"ur Theoretische Physik\\
Universit\"at Heidelberg\\
Philosophenweg 16, D-69120 Heidelberg}

\begin{abstract}
Quantum gravity can determine the dependence of gauge couplings in a scalar field, which is related to possible fifth forces and time varying fundamental \qq{constants}.
This prediction is based on the scaling solution of functional flow equations. For momenta below the field-dependent Planck mass the quantum scale invariant standard model emerges as an effective low energy theory. For a small non-zero value of the infrared cutoff scale the time-variation of couplings or apparent violations of the equivalence principle turn out to be negligibly small for the present cosmological epoch, unless some further quantum scale symmetry violation beyond the standard model comes into play. More sizable effects of quantum scale symmetry violation are expected during nucleosynthesis or before. Scaling solutions relate field dependence and dependence on the renormalization scale. We find asymptotically free gauge couplings for the standard model coupled to gravity, while for grand unified models the gauge couplings are asymptotically safe with non-zero values at the ultraviolet fixed point. The scaling solution of asymptotically safe metric quantum gravity yields restrictions for model building, as limiting the number of scalars in grand unified theories.
\end{abstract}

\maketitle

\section{Introduction}\label{sec: I}

The possible dependence of the electromagnetic fine structure constant on a scalar field $\chi$ has been proposed long ago~\cite{JOR1,DIR,JOR2,JOR3}. Such a field-dependence induces a fifth force by exchange of the scalar field~\cite{PSW2}. If the range of the interaction mediated by the \qq{cosmon} $\chi$ is large enough and the atom-cosmon coupling not too small, the scalar-mediated fifth force becomes observable by an apparent violation of the equivalence principle~\cite{PSW2}. Furthermore, in realistic cosmologies a change of the value of $\chi$ with cosmic time can lead to time-varying fundamental constants~\cite{CWTWC,DAPO, DPV}, for a review see ref.~\cite{UZAN}. If the cosmon $\chi$ is associated with the scalar field responsible for dynamical dark energy or quintessence~\cite{CWQ}, the amount of time-variation of fundamental couplings and the apparent violation of the equivalence principle are related~\cite{CWTVVE}. Both originate from the same atom-cosmon coupling.

In view of the important experimental and observational efforts to investigate apparent violations of the equivalence principle, scalar-induced fifth forces and time varying fundamental couplings, a theoretical determination or limitation of the field-dependence of fundamental couplings would be most welcome. In this note we propose that the scaling solution of quantum gravity indeed fixes the field-dependence of the gauge couplings. We present a first quantitative computation of the strength of the fifth force and variation of the fine structure constant from a fundamental theory, both for the present cosmological epoch and for nucleosynthesis. It fixes similarly the field-dependence of the other dimensionless couplings of the standard model of particle physics, as Yukawa couplings or the quartic Higgs coupling. The present note focuses, however, on the gauge couplings.

For the scaling solution the dimensionless gauge couplings depend only on the ratio of a scalar field $\chi$ over the renormalization scale $k$. The limit $\chi\to0$ corresponds to $k\to\infty$ and explores the behavior near the ultraviolet fixed point. We find that the gauge couplings are asymptotically free for the standard model coupled to quantum gravity, while they are asymptotically safe for grand unified theories. For the scaling solution all relevant parameters at the ultraviolet fixed point vanish. These relevant parameters describe the evolution of small deviations from the scaling solution. The differential equations defining the dependence of couplings on $\chi/k$ for the scaling solution are identical to those describing the $k$-dependence at $\chi=0$ for the more general flow. Our computation can therefore be used directly to infer the critical exponents for the gauge couplings.

In sect.~\ref{sec: QSSM} we establish the close connection between the scaling solution and the quantum scale invariant standard model which emerges as an effective theory for momenta below the $\chi$-dependent Planck mass. Sect.~\ref{sec: RCFF} discusses the relation between field-dependent couplings and a possible fifth force and time variation of fundamental couplings. In sect.~\ref{sec: QGP} we turn to the question if quantum gravity can predict the value of gauge couplings. This depends largely on the sign of the gravity-induced anomalous dimension of the gauge coupling. In sect.~\ref{sec: FFE} we will set up the functional flow equation for the gauge couplings in quantum gravity. The anomalous dimension is discussed in sect.~\ref{sec: ADGC}. Sect.~\ref{sec: SMGU} turns specifically to the standard model and grand unified theories, and we draw conclusions in sect.~\ref{sec: C}.

\section{Quantum scale invariant standard model}\label{sec: QSSM}

The scaling solution of quantum gravity is closely related to quantum scale symmetry~\cite{CWQSS} and we address this topic first. Scale symmetry is realized for theories that do not involve any parameters with dimension of mass or length. Scale symmetry of the classical action for the standard model coupled to gravity is achieved by replacing both the Planck mass and the Higgs boson mass by a scalar field $\chi$~\cite{FUJ1,ETG,PMIN,ZEE,ADL,LSMO,FUJ2},
\bel{I1}
S=\int_x\sqrt{g}\Big\{-\frac{\chi\2}2R+\frac{\lambdh}2\gl h\herm h-\eps\chi\2\gr\2+\dots\Big\}\ ,
\ee
with $h$ the Higgs-doublet, $R$ the curvature scalar and $\sqrt{g}$ the root of the determinant of the metric. The couplings $\lambdh$ and $\eps$ are dimensionless. The dots denote kinetic terms and Yukawa interactions that involve the dimensionless gauge- and Yukawa-couplings.

\subsection*{Quantum scale symmetry}

Often the scale symmetry of the classical action is considered as an approximation, while quantum fluctuations are supposed to lead to a violation of this symmetry. In contrast, for the quantum scale invariant standard model~\cite{CWQ, SHAZEN1} it is the quantum effective action that does not involve any parameter with dimension of mass or length. In this case quantum scale symmetry is an exact global symmetry. If it is broken spontaneously by a non-zero value of the scalar field $\chi$ one predicts the presence of a massless Goldstone boson. Quantum scale symmetry can actually be induced by quantum fluctuations, both for classical actions with and without scale symmetry. Quantum fluctuations generate a flow of couplings or more generally, a flow of coupling functions. Quantum scale symmetry emerges as an exact symmetry for fixed points in the flow of couplings. Those can be related to scaling solutions for couplings functions.

Due to quantum fluctuations the dimensionless couplings become running couplings. The solution of the flow equations for these couplings introduces mass scales as the renormalization scale $\mu$ or the confinement scale $\Lambda\subt{QCD}$ in quantum chromodynamics (QCD). If these mass scales are intrinsic parameters, the quantum effects violate scale symmetry. An example is the running electromagnetic coupling in quantum electrodynamics (QED) for electrons coupled to photons. The running gauge coupling is reflected in the gauge invariant quantum effective action by a gauge invariant kinetic term
\bel{I2}
\Gamma_F=\frac14\int_x\sqrt{g}F\mnb Z\gl-\partial\2\gr F\mno\ ,\quad F\mnb=\partial\mb A\nb-\partial\nb A\mb\ ,
\ee
with $\partial^2=\partial\mo\partial\mb$. In one loop order one has the approximate form
\bel{I3}
Z=\frac1{4\pi\alpha_0}-\frac1{12\pi\2}\ln\left(\frac{-\partial^2+4m\subt{e}^2}{4m\subt{e}^2}\right)\ ,
\ee
with $\alpha_0$ the fine structure constant as measured with experiments for which the relevant length scales are much larger than $m\subt{e}^{-1}$.

In the presence of scale symmetry the electron mass is field-dependent,
\bel{I4}
m\subt{e}=y\subt{e}h_0=y\subt{e}\sqrt{\eps}\chi=h\subt{e}\chi\ ,
\ee
with $h_0$ the expectation value of the neutral component of the Higgs doublet and $y_e$ the electron Yukawa coupling. The dimensionless parameter $h_e=y_e\sqrt{\eps}$ accounts for the exact proportionality between electron mass and the scalar field $\chi$. Inserting eq.~\eqref{I4} into eq.~\eqref{I3} no intrinsic scale appears in $Z$ if $\alpha_0$ and $h_e$ are constants. The effective action~\eqref{I2} is compatible with quantum scale symmetry.

The running of the gauge coupling appears in the coupling of $A_\mu$ to the electrons, which involves the gauge covariant derivative $D_\mu$. Eq.~\eqref{I2} uses a normalization of the gauge field for which the covariant derivative $D\mb=\partial\mb-iA\mb$ does not directly involve gauge coupling. Rescaling the gauge field by $A\mb\to Z^{-1/2}A\mb$ shifts the gauge coupling $g$ to the covariant derivative, and one infers a momentum dependent gauge coupling ($-\partial^2\to q^2$)
\bel{I5}
g^2(q)=Z^{-1}(q)\ .
\ee
For $	q\2\gg4m\subt{e}\2$ it obeys the usual flow equation
\bel{I6}
q\2\frac\partial{\partial q^2}g^2=-Z^{-2}q^2\frac\partial{\partial q^2}Z=\frac{g^4}{12\pi^2}\ ,
\ee
while for $q^2\ll4m\subt{e}^2$ the flow stops. We could also define the gauge coupling at some arbitrary renormalization scale $\mu$, with $g^2(\mu)=\gbar^2$ taking a fixed value. Consistency with the scale invariant expression~\eqref{I3},\eqref{I4} requires that $\mu$ is taken proportional to $\chi$. Otherwise $\alpha_0$ would depend on $\mu$ for fixed $\gbar^2$ and therefore induce a violation of quantum scale symmetry.

For QCD we may again define the gauge coupling by $g^2(\mu)=\gbar^2$. The quantum scale invariant standard model~\cite{CWQ,SHAZEN1} requires $\mu\sim\chi$. In this case the confinement scale where $g^2(q)$ diverges or turns to non-perturbatively large values is proportional to $\chi$
\bel{I7}
\Lambda\subt{QCD}=\eta\subt{QCD}\chi\ .
\ee
With similar procedures for all dimensionless couplings no parameter with dimension of mass or length is present in the quantum effective action. This realizes the quantum scale invariant standard model. All mass scales are dynamical, given by the value of the cosmon field $\chi$.

\subsection*{Running with fields and momentum}

In the quantum scale invariant standard model the running of couplings with momentum is directly related to their field-dependence. Indeed, the dimensionless couplings have to be field-dependent -- $g^2$ depends both on $q^2$ and on $\chi^2$. Being dimensionless, it is a function of the ratio $q^2/\chi^2$. Only for this specific $\chi$-dependence of the dimensionless couplings quantum scale symmetry becomes an exact global symmetry in the presence of quantum fluctuations. For any non-zero value of $\chi$ this global scale symmetry or dilatation symmetry is broken spontaneously. One then predicts an exactly massless Goldstone boson.

For couplings depending only on $q^2/\chi^2$ the renormalization flow with momentum (at fixed $\chi$) is directly linked to the field-dependence of the coupling. This characteristic feature of the quantum scale invariant standard model relates the well known momentum dependence of the couplings to the dependence on scalar fields, and therefore to possible apparent violations of the equivalence principle or to time-varying fundamental constants. Scaling solutions for quantum gravity will introduce an effective infrared cutoff scale $k$. If $\chi/k$ varies in the course of the cosmic evolution, also fundamental parameters can vary. A small scale $k\ll m_e$ (say $k\approx10^{-3}\text{eV}$) will only lead to tiny modifications for the interactions of the particles in the quantum scale invariant standard model. Nevertheless, the Goldstone boson is turned to a pseudo Goldstone boson - the cosmon - with a very small non-zero mass. The interactions mediated by the cosmon need no longer be pure derivative interactions and their effective range is finite.

We observe that quantum scale symmetry does not fix the parameter $\eps$ in eq.~\eqref{I1} and therefore is not related to the gauge hierarchy problem. A possible understanding of the naturalness of the tiny value of $\eps$ in terms of an additional \qq{particle scale symmetry} at fixed effective Planck mass is proposed in refs.~\cite{CWFT,CWMHP,BAR}. This is not the subject of the present work.

\subsection*{Scaling solution in quantum gravity}

Can one get the scale invariant standard model from some consistent quantum field theory that remains valid up to arbitrarily short distances? It has been proposed to employ $\chi$ as an effective ultraviolet cutoff~\cite{CWQ,SHAZEN1,SHAZEN2}. This is valid as long as $q^2\ll\chi^2$, and indeed leads to the quantum scale invariant model. A complete quantum field theory has also to cover the range $q^2>\chi^2$, however, for which the interpretation of $\chi$ as an effective ultraviolet cutoff is no longer valid. If the present value of $\chi$ is associated to the Planck mass $M$, this momentum range concerns transplanckian physics. Similarly, one wants to understand the field-dependence of couplings for the whole range of $\chi$ from zero to infinity. This is particularly relevant for crossover cosmologies for which $\chi$ starts from zero in the infinite past and reaches infinity in the infinite future~\cite{CWVG, CWIQM, CWQGIQ}.

A direct way of obtaining the quantum scale invariant standard model is the scaling solution of quantum gravity. The formulation of quantum gravity as a consistent quantum field theory, which remains valid for all distance scales, requires the existence of an ultraviolet fixed point. Quantum gravity formulated in terms of the metric can be either asymptotically safe~\cite{WEIN1,REU,DOUPER,SOU,LAURE,REUSAU} or asymptotically free~\cite{SWY,STE,FRAV,FRATS}. In turn, such a fixed point needs the existence of scaling solutions for the dependence of dimensionless couplings on $\chi^2/k^2$, with $k$ a suitable renormalization scale. The resulting $\chi$-dependence occurs in addition to the dependence on $q^2/\chi^2$. We will see that for the running gauge couplings the dependence on $k^2/\chi^2$ is closely related to the dependence on $q^2/\chi^2$.

The non-perturbative properties of such a fixed point and the flow towards realistic values can be dealt with by functional renormalization for the effective average action~\cite{CWRG}. This approach introduces an infrared cutoff scale $k$ such that only fluctuations with momenta $q^2>k^2$ are included for the computation of the effective action. Instead of a finite number of couplings one follows the flow of whole functions of fields, which amounts to infinitely many couplings. In particular, the existence of an ultraviolet fixed point requires that the functional flow equations admit a scaling solution for which all dimensionless couplings as the gauge couplings depend only on dimensionless ratios as
\bel{I8}
\tilde\rho=\frac{\chi^2}{k^2}\ ,\quad \qtil^2=\frac{q^2}{\chi^2}\ ,
\ee
without an additional explicit dependence on $k$.

If for $k\to0$ the gauge coupling $g^2(\tilde\rho, \qtil^2)$ reaches a fixed function $g_*^2(\qtil^2)$, any dependence on the scale $k$ disappears. This results in dimensionless momentum-dependent and field-dependent couplings, as for eq.~\eqref{I3} for QED,
\bel{I9}
g^{-2}(\qtil^2)=\frac1{4\pi\alpha_0}-\frac1{12\pi^2}\ln\left(\frac{\qtil^2+4h\subt{e}^2}{4h\subt{e}^2}\right)\ .
\ee
Similarly, one obtains for QCD with $\gbar^2=g^2(q^2=\chi^2)$ and $N\subt{f}$ flavors of massless quarks
\begin{align}
\label{I10}
g^{-2}(\qtil^2)=&\frac1{\gbar^2}+\frac{B_F}{2}\ln(\qtil^2)\ ,\nn\\
B_F=&\frac1{16\pi^2}\gl33-\frac{8N\subt{f}}3\gr\ .
\end{align}
The confinement scale can be identified with the momentum for which the one-loop approximation to the running gauge coupling diverges
\bel{I11}
\Lambda\subt{QCD}^2=\chi^2\exp\left(-\frac2{B_F\gbar^2}\right)\ ,
\ee
realizing eq.~\eqref{I7}. We conclude that the limit $k\to0$ of the scaling solution realizes the quantum scale invariant standard model.

\section{Running couplings and fifth force}\label{sec: RCFF}

The possibility of an observable time variation of fundamental couplings or a scalar fifth force mediating an apparent violation of the equivalence principle results from the interplay of the dependence of dimensionless couplings on $\rhotil$ and $\qtil^2$. For $k=0$ the scalar field $\chi$ describes an exact Goldstone boson. It has only derivative couplings, suppressed by inverse powers of the effective Planck mass $\chi$. The resulting \qq{fifth force effects} are unobservably small. Furthermore, an exact Goldstone boson $\chi$ settles to a constant value very early in cosmology, such that no observable time variation of fundamental couplings occurs due to their dependence on $q^2/\chi^2$. An observable time variation or non-derivative fifth force can therefore only arise for $k\neq0$. (The status of the quantum scale invariant standard model with unimodular gravity~\cite{SHAZEN1} is not clear in this respect. An effective cosmological constant arises there as an integration constant. The dynamics is the same, however, as for an effective scalar potential $\sim k^4$, which violates quantum scale symmetry.)

\subsection*{Running couplings}

We will next establish that for $\rhotil=\chi^2/k^2\gg1$ the dependence of the gauge couplings on $\rhotil$ according to the scaling solution is given by the same perturbative $\beta$-functions as for the usual momentum dependence. These $\beta$-functions do not depend on the details of quantum gravity. Quantum gravity is only needed for the existence of the scaling solution. For large $\rhotil$ and small $\qtil^2$ both the running of the gauge couplings with $\rhotil$ and with $\qtil^2$ only involves the fluctuations of an effective particle theory below the Planck mass. This is typically the standard model of particle physics. If beyond standard model effects, for example in the sector of neutrino masses, can be omitted the fifth force mediated by the cosmon-atom coupling becomes calculable.

Within functional renormalization the flow equations for the $k$-dependence of coupling functions $g^2(q^2,\chi^2,k^2)$ are evaluated at fixed $q^2$ and $\chi^2$
\bel{I12}
\partial_tg^{-2}=k\partial_kg^{-2}\big|_{\chi^2,\qtil^2}=\beta_{g^{-2}}\ .
\ee
For small $g^2$ and $k$ sufficiently below the Planck mass $M$ the $\beta$-function for QED can be approximated by the one-loop expression
\bel{I13}
\beta_{g^{-2}}=B_F\,l\left(\frac{q^2}{k^2},\frac{m\subt{e}^2}{k^2}\right)\ ,\quad B_F=-\frac1{6\pi^2}\ .
\ee
The \qq{threshold function} $l$ accounts for the stop of the flow for $k^2\ll q^2$ or $k^2\ll m\subt{e}^2$, since both $q^2$ and $m\subt{e}^2$ constitute effective infrared cutoffs such that the additional cutoff $~\sim k^2$ no longer matters. We may approximate this decoupling of \qq{heavy modes} from the flow by
\bel{I14}
l=\frac{k^2}{k^2+m\subt{e}^2+q^2/4}\ .
\ee
The solution
\bel{I15}
g^{-2}=g_0^{-2}+\frac{B_F}2\ln\left(\frac{k^2+m\subt{e}^2+q^2/4}{m\subt{e}^2}\right)
\ee
recovers eq.~\eqref{I3} for $k\to0$. It is one of the advantages of functional renormalization that threshold functions as $l$ are produced automatically and extend directly to $\chi$-dependent masses as $m\subt{e}=h\subt{e}\chi$.

In the standard model the non-zero masses cut the flow of all couplings except for non-renormalizable couplings related to the masses of neutrinos. The flow of all renormalizable couplings effectively stops for $k$ below the electron mass. As a result, there is only a negligibly tiny difference between the limit $k\to0$ and some small finite value, say $k\approx 10^{-3}\text{eV}$. Rather realistic cosmology for both inflation and dynamical dark energy obtains for a value of $k$ in the $10^{-3}\text{eV}$ range~\cite{CWQGIQ}. We will take in the following this value. Since $k$ is the only scale appearing in the scaling solution its value is actually arbitrary, defining the mass units. Only the dimensionless ratio $\tilde\rho=\chi^2/k^2$ matters. Our units are chosen such that for a present ratio $\tilde\rho=10^{60}$ one has $\chi=2.44\cdot 10^{18}\text{GeV}$. In runaway cosmologies with ever increasing $\tilde\rho$ the huge value of $\tilde\rho$ is simply an effect of the large age of the universe in Planck units.

From the flow equation~\eqref{I12} at fixed $\chi$ we can switch to the equivalent equation at fixed $\tilde\rho$
\bel{I16}
\gl\partial_t-2\tilde\rho\partial_{\tilde\rho}\gr g^{-2}\big|_{\tilde\rho,\qtil^2}=\beta_{g^{-2}}\ .
\ee
The condition for the scaling solution is that the explicit $k$-dependence at fixed $\tilde\rho$ vanishes. The condition $\partial_t g^{-2}|_{\rhotil,\qtil^2}=0$ yields the defining differential equation for the scaling solution.
\bel{I17}
\tilde\rho\partial_{\tilde\rho}g^{-2}(\tilde\rho)=-\frac12\beta_{g^{-2}}\ .
\ee
For $\rhotil\gg1$, $\qtil^2\ll1$ it only involves the standard model perturbative $\beta$-function.

In the following we mainly consider the limit $\qtil^2\to0$. The scaling solution for the field-dependent coupling $g^2(\tilde\rho)$ has therefore to obey the non-linear differential equation~\eqref{I17} at $\qtil^2=0$. Similar \qq{scaling equations} have to be obeyed for a large coupled system of coupling functions. Away from thresholds, the flow generators ($\beta$-functions) depend only on the dimensionless couplings of the particles that are effectively massless in the corresponding range of $\rhotil$. Threshold functions similar to eq.~\eqref{I14} account for an effective decoupling if $k$ is smaller than the effective particle mass. For the example of the electron the decoupling takes place at $k^2<h_e^2\chi^2$ or $\rhotil>h_e^{-2}$.

As common for such systems of differential equations only few solutions may exist for the whole range $0\leq\tilde\rho<\infty$. It is at this point where quantum gravity can become selective, as we will discuss in the second part of this note. Nevertheless, for the flow of the scaling solution in the range $\rhotil\gg1$ quantum gravity plays no role. Only the existence of the scaling solution is supposed.

\subsection*{Fundamental scale invariance}

Fundamental scale invariance~\cite{CWFSI} postulates that the world is described precisely by a scaling solution, with all relevant parameters for a flow away from the scaling solution set to zero. The requirement of existence of the scaling solution can make this concept rather selective. The absence of relevant parameters makes this scenario highly predictive. 

A theory with fundamental scale invariance can be described in terms of scale invariant fields such that the scale~$k$ never appears. This holds including the continuum limit. In our case the scale invariant fields are $\tilde{\chi}=\chi/ k$ and $\tilde{g}_{\mu\nu}=k^{2}g_{\mu\nu}$.

We may alternatively consider a scenario where the first substantial flow away from the scaling solution would occur at some scale $k_0$. There will be not much difference to the exact scaling solution if we set for the latter $k=k_0$. Evaluating the exact scaling solution at a fixed $k$, say $k=2\cdot10^{-3}\text{eV}$, can alternatively be interpreted as a deviation from the scaling solution at $k_0=2\cdot10^{-3}\text{eV}$ due to a relevant parameter. In both cases $k$ or $k_0$ only fix the units.

As compared to the scaling solution the relevant parameters can add to quantities with dimension $\text{mass}^{N}$ a contribution $\sim k_{c}^{N}$. For the gauge couplings the effect of the running from $k_{c}$ to zero can be respected.

\subsection*{Time variation of couplings and fifth force}

With $\chi$-dependent particle masses as $m\subt{e}=h\subt{e}\chi$ and a $\chi$-dependent Planck mass $M=\chi$ one may expect the presence of a fifth force mediated by cosmon exchange. The issue for observational consequences is more subtle, however, since the mass ratio $m\subt{e}/M=h\subt{e}$ is independent of $\chi$. For a discussion of observations it is best to make a Weyl transformation to the Einstein frame. In the Einstein frame the Planck mass $M=\Mbar$ is a constant which is introduced only by the variable transformation of the metric $g\mnb'=\gl\chi^2/\Mbar^2\gr g\mnb$. Also the particle masses are constant, $m\subt{e}=h\subt{e}\Mbar$, $\Lambda\subt{QCD}=\eta\subt{QCD}\Mbar$. Dimensionless couplings as $h\subt{e}$ or $g$ remain invariant under a Weyl scaling. The ratio $\qtil^2$ does not change its value and is given in the Einstein frame by $\qtil^2=q^2\subt{E}/\Mbar^2$. (Note $q^2\subt{E}=q_\mu q_\nu {g'}\mno=\gl\Mbar^2/\chi^2\gr q^2=\Mbar^2\qtil^2$.)

Writing eq.~\eqref{I15} in the form
\begin{align}
\label{I18}
g^{-2}=&g_0^{-2}+\frac{B_F}2\ln\left(1+\frac1{h\subt{e}^2\tilde\rho}+\frac{\qtil^2}{4h\subt{e}^2}\right)\nn\\
=&g_0^{-2}+\frac{B_F}2\ln\left(1+\frac{\Mbar^2k^2}{m\subt{e}^2\chi^2}+\frac{q\subt{E}^2}{4m\subt{e}^2}\right)\ ,
\end{align}
a dependence on $\chi$ remains only for $k\neq0$. (The second equation uses the fixed value $m\subt{e}=h\subt{e}\Mbar$.) For the exact quantum scale invariant standard model for $k=0$ only the derivative couplings of a Goldstone boson remain.

For a present value of $\chi$ equal to $\Mbar$ the cosmon coupling to atoms is suppressed by a tiny factor $k^2/m\subt{e}^2\approx 10^{-17}$. The quantum scale invariant standard model is a very good approximation, predicting the absence of non-derivative atom cosmon couplings. Our scenario predicts with high precision the absence of an apparent violation of the equivalence principle or a time variation of fundamental couplings in the present cosmological epoch. The only loophole could be a violation of quantum scale symmetry in the beyond standard model sector, related to an effective dependence of the ratio neutrino mass over electron mass, $m_\nu/m_e$, on $\rhotil$~\cite{CWQSS}.

In earlier epochs of cosmology when $\tilde\rho$ assumed smaller values the time variation of $g^2$ with varying $\tilde\rho$ and the corresponding fifth force are more substantial. This happens for $\tilde\rho<h\subt{e}^{-2}$~\cite{CWQGIQ}. In this range the difference of the fine structure constant from the present value is given according to eq.~\eqref{I18} by
\begin{align}
\label{18A}
\Delta\alpha&=\alpha(\rhotil)-\alpha_0=\Big\{\big[1+\frac{\alpha_0}{3\pi}\ln\gl h_e^2\rhotil\gr\big]^{-1}-1\Big\}\alpha_0\nn\\
&\approx-\frac{\alpha_0^2}{3\pi}\ln\gl h_e^2\rhotil\gr\ .
\end{align}

For an estimate of $\rhotil$ in the radiation dominated cosmological epoch we observe that according to the scaling solution the scalar potential divided by the fourth power of the effective Planck mass obeys for $\chi\to\infty$~\cite{PRWY}
\bel{18B}
\frac{U}{M^4}=\frac{u_\infty k^4}{\chi^4}\ ,\quad u_\infty=\frac {5}{256\pi^2}\ .
\ee
This ratio is the same in all metric frames. 
Eq.~\eqref{18B} determines the order of magnitude also for finite large $\chi/k$ such that we can employ eq.~\eqref{18B} replacing $\mu_{\infty}\to d_{u}\mu_{\infty}$ with $d_{u}$ of the order one. This also includes the case of a relevant parameter which adds to $U/M^{4}$ a part $\lambda k^{4}/\chi^{4}$ with $\lambda$ of the order $u_{\infty}\,$.

Let us assume that in the radiation dominated era cosmology one has a small fraction $\Omega_e$ of early dark energy. This implies in the Einstein frame
\bel{18C}
U=f_{e}\Omega_e\rho_c=3f_{e}\Omega_{e}M^{2}H^{2}=f_{e}\Omega_{e}c_{T} T^4\ ,
\ee
with $f_{e}=(1-w_{e})/2$ involving the equation of state $w_{e}$ of early dark energy. For a cosmic scaling solution where dark energy assumes a constant fraction of the radiation energy density one has $f_{e}=1/3$. If the kinetic energy of the scalar is negligible one finds $f_{e}=1$.  The critical energy density $\rho_c$ is related to the temperature $T$ by $\rho_c=c_TT^4$. Eq.~\eqref{18B} results in
\bel{18D}
\rhotil=\left(\frac{d_{u}u_\infty}{f_{e}\Omega_{e}c_T}\right)^{\frac12}\frac{M^2}{T^2}\ .
\ee
For large enough $T$ eq.~\eqref{18A} implies a relative change of the fine structure constant
\bel{18E}
\frac{\Delta\alpha}{\alpha}=\frac{2\alpha}{3\pi}\bigg[\ln\left(\frac{T}{m_e}\right)+\frac14\ln\left(\frac{f_{e}\Omega_{e}c_T}{d_{u}u_\infty}\right)\bigg]\ .
\ee
This formula becomes valid if the square bracket exceeds one. In this range $\Delta\alpha/\alpha$ exceeds a value $2\alpha/(3\pi)\approx1.55\cdot10^{-3}$. For smaller $T$ the relative change of $\alpha$ is reduced according to
\begin{align}
\label{18F}
\frac{\Delta\alpha}{\alpha}&=\frac{\alpha}{3\pi}\ln\left(1+\sqrt{\frac{f_{e}\Omega_{e}c_T}{d_{u}u_\infty}}\frac{T^2}{m_e^2}\right)\nn\\
&\approx7.7\cdot10^{-4}\ln\left(1+12.9-\sqrt{\frac{f_{e}\Omega_{e}g\subt{eff}}{d_{u}}}\frac{T^2}{m_e^2}\right)\ .
\end{align}
Here we employ eq.~\eqref{18B} for $u_\infty$ and $c_T=(\pi^2/30)g\subt{eff}$, with $g\subt{eff}=29/4$ for $T\lesssim m_e$ and $g\subt{eff}=43/4$ for $T\gtrsim m_e$. To our knowledge eq.~\eqref{18F} is the first example of a quantitative estimate of the variation of the fine structure constant from a well defined setting of a fundamental theory.

Nucleosynthesis occurs in a range where $T/m_e\approx0.2$. For $f_{e}/d_{u}\approx 1$ this yields in the range relevant for nucleosynthesis the approximate value ($g\subt{eff}=29/4$)
\bel{18G}
\frac{\Delta\alpha}{\alpha}\approx 10^{-3}\sqrt{\Omega_e}\left(\frac{5T}{m_e}\right)^2\ .
\ee
There are severe upper bounds on $\Omega_{e}\ $. For a cosmic scaling solution one expects $\Omega_{e} $ to be similar to later stages of radiation domination, where one finds $\Omega_{e}<10^{-2}$~\cite{Gomez-Valent:2021cbe}. Direct bounds from the effective number of massless particles during nucleosynthesis are similar, albeit slightly weaker. If the scalar field has settled to a value for which the potential becomes relevant only after nucleosynthesis the value of $\Omega_{e}$ is smaller than for the cosmic scaling solution. We conclude $\Delta\alpha/\alpha\lesssim 10^{-4}$, unless a kination epoch ends only very close to nucleosynthesis.

A variation of the fine structure constant affects the relative element abundances $Y_i$ obtained from primordial nucleosynthesis according to
\bel{18H}
\frac{\Delta Y_i}{Y_i}=c_i\frac{\Delta\alpha}{\alpha}\ ,
\ee
with $c_i=(3.6,1.9,-11)$ for $i=(\text{D},{}^4\text{He},{}^7\text{Li})$~\cite{DSW}. A variation $\Delta\alpha/\alpha\lesssim 10^{-4}$ seems to be too small for an observable modification of the primordial element abundances. Due to the strong dependence of $\Delta\alpha/\alpha$ on $T$ a more detailed quantitative estimate would need to follow explicitly the time-variation of $\alpha$ during nucleosynthesis. Similar considerations can be made for the field dependence of other couplings, as the Yukawa coupling of the electron, which affects the ratio of electron to proton mass.

\section{Quantum gravity predictions}\label{sec: QGP}

In the second part of this note we investigate the question if a scaling solution for the gauge couplings exists and what are its properties. This necessarily involves the effects of the metric fluctuations in quantum gravity. We need to compute the flow equations for gauge couplings in quantum gravity. Their detailed form will decide if the gauge couplings are asymptotically free or safe, and if quantum gravity can make a prediction for the observed values of the gauge couplings. This second part contains the main technical advances of this work.

\subsection*{Ultraviolet completion and quantum gravity}

So far we have assumed the existence of the scaling solution for $\rhotil>\rhotil\subt{c}$, with $\rhotil\subt{c}\gg1$ such that the metric fluctuations in quantum gravity can be neglected. The question is what happens for $\rhotil<\rhotil\subt{c}$? The answer needs an extension of the flow equation to quantum gravity. Functional renormalization has found for the flow of the gauge coupling without a scalar field a gravity-induced anomalous dimension $B_g$~\cite{DHR1, Folkerts:2011jz, Harst:2011zx, Christiansen:2017gtg, CLPR, EV, DHR2, deBrito:2022vbr, Eichhorn:2022vgp} see also~\cite{Robinson:2005fj, Pietrykowski:2006xy, Toms:2007sk, Ebert:2007gf, Tang:2008ah, Toms:2010vy},
\bel{I19}
\partial_tg^2=-B_gg^2-B_Fg^4\ ,
\ee
with
\bel{I20}
B_g=2C_g\frac{k^2}{M^2(k)}=\frac{C_g}{w}\ ,\quad w=\frac{M^2(k)}{2k^2}\ .
\ee
More details will be presented in later parts of this note.
For the UV-fixed point $w$ approaches a constant $w_*$ and therefore $B_g$ becomes constant. On the other hand, the gravitational contribution decreases rapidly for $k^2\ll M^2(k)$.

The flow equation~\eqref{I19} is easily extended to the presence of a scalar field by having $M^2$ depending on $\chi^2$ and on $k$. A typical form of the scaling solution for $w$ is given by~\cite{HPRW, HPW, WY}
\bel{I20*}
w=w_0+\frac12\rhotil\ ,
\ee
where we have taken a normalization of the scalar field such that for $\chi\to\infty$ one has $M^2=\chi^2$. According to eq.~\eqref{I17} the scaling solution for the gauge coupling has to obey
\bel{I21}
\rhotil\partial_{\rhotil}g^2=\frac{C_gg^2}{2w_0+\rhotil}+\frac{B_Fg^4}2\ .
\ee
The UV-limit corresponds to $\rhotil\to0$, or $k\to\infty$ at fixed $\chi$. For the UV-completion of the scaling solution we therefore need to consider the region where $\rhotil$ can be neglected in the denominator of the first term in eq.~\eqref{I21}. For this regime both $B_g=C_g/w_0$ and $B_F$ are constant. The constant $B_F$ arises from the fluctuations of the gauge bosons and charged particles. It includes contributions from all particles that are effectively massless in the transplanckian regime for $\rhotil\to0$. Typically one expects massless particles beyond the ones present in the standard model as an effective low energy theory.

Depending on the sign of $B_g$ and $B_F$ we distinguish four scenarios:
\begin{enumerate}[(i)]

\item\label{list1i} $B_g>0$, $B_F>0$: In this case the gauge coupling is asymptotically free. It reaches zero for $\rhotil\to0$. A family of scaling solutions can be characterized by the value
\bel{I21*}
g^2(\rhotil_d)=g^2_d\ ,\quad \rhotil_d\ll1\ .
\ee
Here $\rhotil_d$ replaces the usual renormalization scale in case of a scaling solution. Its value can be chosen freely as long as $\rhotil_d\ll1$.
For increasing $\rhotil>\rhotil_d$ the gauge coupling increases. For the gauge bosons of the standard model the value of $g_d^2$ has to be small enough such that at the decoupling of the gravity fluctuations for $\rhotil\approx2w_0$ the gauge couplings are not too large. Observation requires $g^2(\rhotil=20w_0)\approx0.3$. If $g_d^2$ remains below the resulting upper bound, the flow can be continued from $\rhotil=20w_0$ towards large $\rhotil$, matching the discussion above for $\rhotil\gg2w_0$ where only the part $\sim B_F$ contributes to the flow.

At this level there are continuous families of scaling solutions parametrized by different values of $g_d^2$. Obtaining phenomenologically acceptable values of $g^2(\rhotil=20w_0)$ does not require relevant parameters for deviations from the scaling solution. One rather has to pick one out of the family of scaling solutions. What is not guaranteed, however, is that all members of the family of scaling solutions result in a viable scaling solution for the coupled system of many coupling functions.

\item\label{list1ii} $B_g>0$, $B_F<0$: In this case the gauge coupling remains asymptotically free. Its flow exhibits now an infrared stable fixed point
\bel{I22}
g_*^2=-\frac{B_g}{B_F}\ .
\ee
For $g_d^2<g_*^2$ the gauge coupling $g^2$ increases for increasing $\rhotil$, approaching rapidly the value $g_*^2$. On the other hand, for $g_d^2>g_*^2$ the gauge coupling decreases towards $g_*^2$. For a consistent scaling solution we should be able to choose $\rhotil_d$ arbitrarily close to one. This implies a bound
\bel{I23}
0\leq g^2(\rhotil=2w_0)\lesssim g_*^2\ .
\ee
For small enough $\rhotil_d$ arbitrarily large values of $g_d^2$ are mapped at $\rhotil=2w_0$ to the fixed point value $g^2(2w_0)\approx g_*^2$. This follows from the dominance of the term $B_F$ for large $g^2$, according to
\bel{I24}
\rhotil\partial_{\rhotil} g^{-2}=B_gg^{-2}+B_F\ .
\ee
The restricted range~\eqref{I23} is a first example for the predictivity of scaling solutions. The infrared stable fixed point for $B_g>0$, $B_F<0$ has been discussed for abelian gauge theories in ref.~\cite{DHR1}, and for grand unified theories in refs.~\cite{EHW1,EHW2}.

\item\label{list1iii} $B_g<0$, $B_F>0$: For this case the gauge coupling is asymptotically safe instead of asymptotically free. At the UV-fixed point the gauge couplings differ from zero, taking the value~\eqref{I22}. The fixed point~\eqref{I22} becomes infrared unstable. Unless $g_d^2$ is chosen very close to $g_*^2$ the coupling $g^2(\rhotil=2w_0)$ is either zero or it diverges before $\rhotil=2w_0$ is reached. Both cases are unacceptable. Only the possibility of tuning $g_d^2$ very close to $g_*^2$ remains for very small $\rhotil_d$.

\item\label{list1iv} $B_g<0$, $B_F<0$: For this situation the theory is trivial in the gauge sector. Any positive coupling $g_d^{-2}\geq0$ at $\rhotil_d\to0$ is mapped to $g^2(\rhotil=2w_0)=0$. This prediction of the scaling solution is not compatible with observation.

\end{enumerate}

\subsection*{Predictivity}

From the point of view of consistency the requirement of existence of the scaling solution leads to predictions in two cases: $g^2(2w_0)<g_*^2$ for case~(\ref{list1ii}), and $g^2(2w_0)=0$ for case~(\ref{list1iv}). Further predictivity is gained if we insist on avoiding tuning of parameters. Assume at $\rhotil_d\ll1$ some arbitrary value $g_d^2$ not extremely close to zero and not extremely close to $g_*^2$. For a sizable $B_g$ this value changes quickly with $\rhotil$. For small $g^2$ it obeys
\bel{I25}
g^2(\rhotil)=g_d^2\left(\frac{\rhotil}{\rhotil_d}\right)^{B_g/2}\ ,
\ee
while in the presence of a second fixed point $g_*^2$ one has for small $g^2-g_*^2$
\bel{I26}
g^2(\rhotil)=g_*^2+\gl g_d^2-g_*^2\gr\left(\frac{\rhotil}{\rhotil_d}\right)^{-B_g/2}\ .
\ee
Any value of $g_d^2$ away from the fixed point is driven rapidly towards the infrared stable fixed point, $g_*^2=-B_g/B_F$ for case~(\ref{list1ii}) or $g_*^2=0$ for the cases~(\ref{list1iii}) and~(\ref{list1iv}). For case~(\ref{list1i}), as well as for the case~(\ref{list1iii}) with $g_d^2>g_*^2$, the gauge coupling diverges before $\rhotil$ reaches the value $2w_0$.

We conclude that a particularly robust situation arises for case~(\ref{list1iii}) with a predicted value
\bel{I27}
g^2(\rhotil=2w_0)=-\frac{B_g}{B_F}\ .
\ee
Comparison of this prediction with the value $g^2\approx0.3$ compatible with observation requires knowledge of $B_g$ and $B_F$. The constant $B_F$ can be determined by the usual one-loop formula in the absence of gravity. It involves the number of massless gauge bosons, fermions and scalars for transplanckian physics. If $B_g$ is computed reliably, compatibility with observation restricts the particle content of transplanckian physics.

Previous computations have mainly found positive $B_g$. The issue of the sign is complex, however, since two diagrams contribute with opposite sign. Typically, one also finds parameter ranges for which $B_g$ is negative. The results obtained so far show substantial differences both for the absolute magnitude of $B_g$ and the boundary in parameter space separating positive and negative $B_g$. In view of the importance of $B_g$ for restricting the transplanckian particle content we compute this quantity in the next part of this note by employing the gauge invariant formulation of the functional flow equation~\cite{CWGIFE}. This formulation separates clearly between physical fluctuations and gauge fluctuations. Furthermore, the contribution of the gauge fluctuations combined with the regularized Faddeev-Popov determinant or ghost sector yields a universal measure contribution which does not depend on the form of the effective average action $\Gamma_k$. A clear separation of the different modes provides for a simple physical picture of the flow and is expected to be particularly robust.

\section{Functional flow equations}\label{sec: FFE}

We start with the simplest truncation of the gauge invariant effective average action $\Gamma_{k}$,
\bel{1}
\Gamma_{k}=\int_{x}\sqrt{g}\Big{\lbrace} \frac{Z(\chi)}{4}F_{\mu\nu}^{z}F_{z}^{\mu\nu}-\frac{1}{2}M\2(\chi)R+V(\chi)+\mathcal{L}\subt{kin}[\chi ] \Big{\rbrace}\ .
\ee
Here $Z(\chi)$, $M\2(\chi)=2w(\chi)k\2$ and $V(\chi)=u(\chi)k^{4}$ depend on the scalar field $\chi$, and $\mathcal{L}\subt{kin}[\chi]$ is a kinetic term for $\chi$. For a scaling solution the dimensionless functions $Z$, $w$ and $u$ depend only on the dimensionless ratio $\rhotil=\chi\2/k\2$. The non-abelian field strength,
\bel{2}
F_{\mu\nu}^{z}=\partial_{\mu}A_{\nu}^{z}-\partial_{\nu}A_{\mu}^{z}+ f_{uv}{}^{z}A_{\mu}	^{u}A_{\nu}^{v}\ ,
\ee
involves the structure constants $f_{uv}{}^{z}$ of the gauge group. Indices are raised with the inverse metric $g^{\mu\nu}$ and $g=\det(g_{\mu\nu})$, such that $\sqrt{g}$ includes a factor $i$ in case of Minkowski signature. We are interested in the flow of the gauge coupling $\alpha=g^2/(4\pi)=(4\pi Z)^{-1}$ with $k$ and possible scaling solutions for which $Z$ depends only on $\rhotil$. We focus on constant field values $\chi$ and do not explicitly compute the effect of scalar fluctuations. Thus $\mathcal{L}\subt{kin}[\chi]$ may be neglected.

The truncation~\eqref{1} omits terms involving four derivatives of the metric which are quadratic in the curvature tensor, as $R_{\mu\nu\rho\sigma}R^{\mu\nu\rho\sigma}$ or $R^2$. It is suitable for asymptotically safe gravity defined by the (generalized) Reuter fixed point. It does not cover the short distance behavior of asymptotically free quantum gravity, for which the graviton propagator appearing in the flow equation has to be modified according to ref.~\cite{SWY}. For asymptotically free quantum gravity the computation of this note may still be relevant if there exists an intermediate range of scales $k$ for which the influence of the higher derivative terms can be neglected.

\subsection*{Gauge invariant flow equation}

The exact flow equation\cite{CWRG,RW}, 
\bel{3}
k\partial_{k}\Gamma\kk=\frac{1}{2}\tr\Big{\lbrace}\big{(}\Gamma\kk^{(2)}+R\kk\bigr{)}^{\!-1}k\partial\kk R\kk\Big{\rbrace}-\delta\kk\ ,
\ee
involves the second functional derivative $\Gamma\kk^{(2)}$ of the effective average action, and $R\kk$ is a suitable infrared cutoff function. For the gauge invariant formulation of the flow equation~\cite{CWGIFE} the first term in eq.~\eqref{3} involves only the contributions from the physical fluctuations. The appropriate projection is realized by a particular \qq{physical gauge fixing}, and $R_k$ acts only on the physical fluctuations. In this formulation $\delta\kk$ is a measure contribution with a similar structure as the first term, see below.
The flow of $Z=g^{-2}$ can be extracted by evaluating $\partial\kk\Gamma\kk$ for non-zero gauge fields and a flat space metric.

\subsection*{Field-dependent inverse propagator matrix}

The flow equation depends on the field-dependent propagator for the physical fluctuations in the presence of the cutoff, $(\Gamma_k^{(2)}+R_k)^{-1}$. Since it involves the second functional derivative of the effective average action eq.~\eqref{3} is a functional differential equation. The approximation consists in evaluating $\Gamma_k^{(2)}$ for the particular truncation~\eqref{1}. For the computation of $\Gamma_k^{(2)}$ for field configurations with non-zero gauge fields $\bar{A}_\mu^z$ and flat geometry we expand
\bel{4}
g_{\mn}=\eta_{\mn}+h_{\mn}\;,\quad A\mb^{z}=\bar{A}\mb^{z}+a\mb^{z}\ .
\ee
We work in euclidean space, $\eta_{\mn}=\delta_{\mn}$, and may continue analytically to Minkowski space at the end. We also consider a scalar field $\chi=\overline\chi+\delta\chi$. Since we do not compute here the effect of scalar fluctuations we can omit $\delta\chi$ and simply consider constant $\chi$.

For an evaluation of the inverse propagator $\Gamma_k^{(2)}$ in the presence of the gauge fields $\bar A_\mu^z$ we expand the truncation~\eqref{1} in quadratic order in $a\mb^{z}$ and $h_{\mn}$
\bel{5}
\Gamma_{k,2}=\Gamma_{a}+\Gamma_{h}+\Gamma_{hg}+\Gamma_{F1}+\Gamma_{F2}\ .
\ee
The term quadratic in the gauge field fluctuations
\bel{6}
\Gamma_{a}=\int_{x}\frac{Z}{2}a\mb^{z}(\mathcal{D}_{T})_{zy}^{\mn}a\nb^{y}\ ,
\ee
involves the operator
\bel{7}
(\mathcal{D}_{T})_{zy}^{\mn}=-(D^{\rho}D_{\rho})_{zy}\eta^{ \mn}+(D\mo D\no)_{zy}+2f^{u}{}_{zy}F_{u}^{\mn}\ ,
\ee
with
\bel{8}
D\mb a\nb^{z}=\partial\mb a\nb^{z}+f_{uv}{}^{z}A\mb^{u} a\nb^{v}\ .
\ee
Here and in the following we omit the bar on the ``background gauge fields".

For the terms quadratic in the metric fluctuations one has ($h=h\mb^{\mu}\,$, $\,\partial\2=\partial\mo\partial\mb$)
\bel{9}
\Gamma_{h}=\frac{1}{8}\int_{x}\Big{\lbrace}-M\2(h^{\mn}\partial\2 h_{\mn}-h\partial\2 h)+V(h\2-2h_{\mn}h^{\mn})\Big{\rbrace}\ ,
\ee
and
\bel{10}
\Gamma_{hg}=-\frac{1}{4}\int_{x}M\2(\partial_{\rho}h_{\nu}^{\rho}-\partial_{\nu}h)\partial\mb h^{\mn}\ .
\ee
The first term $\Gamma_h$ concerns the physical metric fluctuations, while the second $\Gamma_{hg}$ involves the gauge fluctuations.

Furthermore, there are contributions $\sim F\2$
\begin{align}
\label{11}
\Gamma_{F2}=\int_{x}&\frac{Z}{4}\Big{[}F_{\mn}^{z}F_{z\rho\sigma}h^{\mu\rho}h^{\nu\sigma}\nn\\
&+F_{\mn}^{z}F_{z\rho}{}^{\nu}(2h_{\tau}^{\mu}h^{\tau\rho}-hh^{\mu\rho})\nn\\
&+\frac{1}{8}F_{\mn}^{z}F_{z}^{\mn}(h\2 -2h_{\rho\tau}h^{\rho\tau})\Big{]}\ .
\end{align}
Finally, $\Gamma_{F1}$ mixes the gauge field and metric fluctuations
\bel{12}
\Gamma_{F1}=\int_{x}\frac{Z}{4}(D\mb a\nb^{z}-D\nb a\mb^{z})(hF_{z}^{\mn}-4h^{\mu\rho}F_{z\rho}{}^{\nu})\ .
\ee

We consider a restricted set of ``background field strengths" $F_{\mn}$ for which the covariant derivative vanishes
\bel{14}
D_{\rho}F_{\mn}^{z}=0\ .
\ee
This simplifies the mixing term, using partial integration
\bel{15}
\Gamma_{F1}=\int_{x}ZF_{z}^{\mu\rho}\Big{(}\partial\mb h_{\rho}\no-\frac{1}{2}\partial\mb h \delta_{\rho}\no \Big{)}a\nb^{z}\ .
\ee
For an extraction of the flow of $Z$ it is sufficient to evaluate the flow of the effective action for constant gauge field strength $F_{z}^{\mn}$. This will further simplify the flow equation.

\subsection*{Physical and gauge fluctuations}

For the  setting of the gauge invariant flow equation~\cite{CWGIFE} we split the metric fluctuations into physical fluctuations and gauge fluctuations~\cite{CWMF}
\bel{AA1}
h_{\mn}=t_{\mn}+\frac{1}{3}\tilde{P}_{\mn}\sigma +a_{\mn}\ .
\ee
In momentum space the physical graviton fluctuation is traceless and transverse, and the physical scalar fluctuation $\sigma$ in the metric is multiplied by a projector $\tilde{P}_{\mn}$, 
\bel{AA2}
q\no t\mnb =0\;,\quad t\mb\mo =0\;,\quad \tilde{P}_{\mn}=\eta\mnb -\frac{q\mb q\nb}{q\2}\ .
\ee
The gauge fluctuations of the metric $a\mnb$ are treated separately. Their contribution, together with the regularized Faddeev-Popov determinant or associated ghost fluctuations, yields the universal measure contribution in the functional flow equation. The same holds for the longitudinal gauge field fluctuations which correspond to the gauge degrees of freedom of the local gauge symmetry. The combined measure contributions are denoted by $\delta\kk$ in eq~\eqref{3}.

The measure contribution for the metric does not depend on the gauge fields. It therefore does not contribute to the flow equation for $Z$. The measure contribution for the gauge fields is the standard one in flat space~\cite{RW, CWG}. The simple separate treatment of the measure contribution is an important technical advantage of the gauge invariant formulation of the functional flow equation. In our truncation an identical result can be obtained in a standard background gauge fixing procedure with ghosts, using a \qq{physical gauge fixing}. For the gauge fields this is Landau gauge, while for the metric the gauge fixing term $(D^\mu h_{\mn})^2$ is multiplied by a coefficient that is taken to infinity. As a result, the propagator matrix becomes block diagonal in the physical and gauge fluctuations. This gauge fixing operates an effective projection of the propagator on the physical degrees of freedom. For the part of the physical fluctuations we can simply impose the ``physical gauge fixing"
\bel{13}
\partial^{\mu}h_{\mn}=0\;,\quad D^{\mu}a\mb^{z}=0\ .
\ee
As a result, $\Gamma_{hg}$ in eq.~\eqref{10} can be omitted.

In the sector of physical fluctuations we can write the matrix $\Gamma_{k}^{(2)}+R_{k}$ in block form
\begin{equation}\label{leer01}
\Gamma_{k}^{(2)}+R_{k}=\begin{pmatrix}
P^{(h)}+P^{(h,F)} & , &P^{(ha)}\\
P^{(ha)\dagger}& , & P^{(a)}
\end{pmatrix}\ .
\end{equation}
In momentum space one has for the physical metric fluctuations
\begin{align}\label{leer02}
P_{\mu\nu\rho\sigma}^{(h)}=\frac{M\2}{4}\bigg{[}&\Big{(}q\2-\frac{2V}{M\2}\Big{)}P_{\mu\nu\rho\sigma}^{(t)}\nn\\
&-2\Big{(}q\2-\frac{V}{2M\2}\Big{)}P_{\mu\nu\rho\sigma}^{(\sigma)}+\tilde{R}_{k,\mu\nu\rho\sigma}^{(h)}\bigg{]}\ ,
\end{align}
where the projectors on the graviton fluctuations and scalar metric fluctuations read
\begin{align}\label{leer03}
\Phat_{\mu\nu\rho\sigma}^{(t)}&=\frac{1}{2}(\tilde{P}_{\mu\rho}\tilde{P}_{\nu\sigma}+\tilde{P}_{\mu\sigma}\tilde{P}_{\nu\rho})-\frac{1}{3}\tilde{P}_{\mn}\tilde{P}_{\rho\sigma}\ ,\nn\\
\Phat_{\mu\nu\rho\sigma}^{(\sigma)}&=\frac{1}{3}\tilde{P}_{\mn}\tilde{P}_{\rho\sigma}\;,\quad \Phat^{(t)}\Phat^{(\sigma)}=0\ .
\end{align}
The projector on the physical metric fluctuations reads
\bel{20A}
\Phat_{\mu\nu\rho\sigma}^{(f)}=\Phat_{\mu\nu\rho\sigma}^{(t)}+\Phat_{\mu\nu\rho\sigma}^{(\sigma)}.
\ee

For the transverse gauge boson fluctuation block one has
\begin{equation}\label{TDA3}
P_{\;\; zy}^{(a)\mn}=Z\Big{\lbrace}(\mathcal{D}_{T})_{zy}^{\mn}+\tilde{R}_{k \,,\;\, zy}^{(a)\, \mn}\Big{\rbrace}\ .
\end{equation}
The off-diagonal term is linear in $F_{\mn}$. For constant $F_{\mn}$ it reads in momentum space
\begin{equation}\label{TDA4}
P_{\, z}^{(ha) \mn , \rho}=-\frac{i}{2}q_{\sigma}ZF_{z}{}^{\sigma \tau}\big{(}\delta_{\tau}\mo\eta^{\nu\rho}+\delta_{\tau}\no\eta^{\mu\rho}-\delta_{\tau}^{\rho}\eta^{\mu\nu}\Big{)}\ .
\end{equation}
Here we have assumed that the infrared cutoff $R_{k}$ is block diagonal. Finally, one has
\begin{align}\label{N1}
P^{(h F)\mn  \rho\sigma}=&\frac{Z}{4}\Big{\lbrace}F_{z}^{\mu\rho}F^{z\nu\sigma}+F_{z}{}^{\mu\sigma}F^{z\nu\rho}\nn\\
&+F_{z}{}^{\mu}{}_{\tau}F^{z\sigma\tau}\eta^{\nu\rho}+F_{z}^{\nu}{}_{\tau}F^{z\sigma\tau}\eta^{\mu\rho}\nn\\
&+F_{z}{}^{\mu}{}_{\tau}F^{z\rho\tau}\eta^{\nu\sigma}+F_{z}{}^{\nu}{}_{\tau}F^{z\rho\tau}\eta^{\mu\sigma}\\
&-F_{z}^{\rho}{}_{\tau}F^{z\sigma\tau}\eta^{\mn}-F_{z}{}^{\mu}{}_{\tau}F^{z\nu\tau}\eta^{\rho\sigma}\nn\\
&+\frac{1}{4}F_{z}{}^{\tau\lambda}{}F^{z\tau\lambda}\big{(}\eta^{\mn}\eta^{\rho\sigma}-\eta^{\mu\rho}\eta^{\nu\sigma}-\eta^{\mu\sigma}\eta^{\nu\rho}\big{)}\Big{\rbrace}\ .\nn
\end{align}

\subsection*{Flow equation for gauge coupling}

The flow equation for $Z$ and therefore the gauge coupling $g^2=Z^{-1}$ can be extracted by evaluating the contribution to the flow equation~\eqref{3} which is quadratic in $F\mnb$. We can therefore expand for small $F\mnb$. Since the off-diagonal term $P^{(ha)}$ is proportional to $F\mnb$ we can use a matrix expansion of the inverse propagator. Writing
\begin{align}\label{N2}
\big{(}\Gamma_{k}^{(2)}+R_{k}\big{)}&=D+M \phantom{\Big{|}}\ ,\\
D=\begin{pmatrix}
P^{(h)}+P^{(h,F)} & 0 \\
0 & P^{(a)}
\end{pmatrix}&\;,\quad M=\begin{pmatrix}
0 & P^{(ha)}\\ P^{(ha)\dagger} & 0 
\end{pmatrix} \ ,\nn
\end{align}
we need the second order in M
\bel{N3}
\big{(}\Gamma_{k}^{(2)}+R_{k}\big{)}^{-1}=D^{-1} - D^{-1} M D^{-1} +D^{-1} M D^{-1} M D^{-1}+\dots
\end{equation}
With block diagonal $R_k$ the term linear in $M$ does not contribute to the trace. The remaining terms are block diagonal and we can invert the inverse propagators for the different blocks separately. The propagators in each block involve the same projections on physical modes as for the inverse propagator.

\begin{figure}[h]
\centering\includegraphics{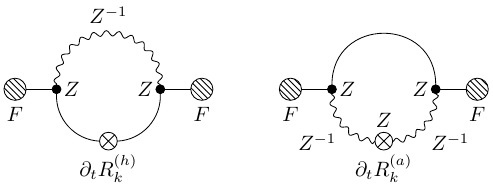}

\vspace*{-5pt}

\caption{Graphical representation of mixed contribution $\pi\kk^{(ah)}$ to the flow of the effective action. Curled lines are gauge boson propagators, and solid lines indicate the propagators for the metric fluctuations. Gauge boson propagators carry a factor $Z^{-1}$, vertices a factor $Z$, and $\partial_tR_k^{(a)}\sim Z$.}\label{Fig.1}
\end{figure}

For block diagonal $R_{k}$ one concludes, with $\partial_{t}=k\partial_{k}$,
\bel{N4}
\pi\kk=\frac{1}{2}\tr \Big{\lbrace}\big{(}\Gamma\kk^{(2)}+R\kk\big{)}^{-1}\partial_{t}R\kk\Big{\rbrace}=\pi\kk^{(a)}+\pi\kk^{(h)}+\pi\kk^{(ha)}\ ,
\ee
with
\begin{align}
\label{N5}
\pi\kk^{(a)}&=\frac{1}{2}\tr\Big{\lbrace}P^{(a)-1}\partial_{t}R\kk^{(a)}\Big{\rbrace}\ ,\nn\\
\pi\kk^{(h)}&=\frac{1}{2}\tr\Big{\lbrace}\big{(}P^{h}+P^{(hF)}\big{)}^{-1}\partial_{t}R\kk^{(h)}\Big{\rbrace}, \nn\\
\pi\kk^{(ah)}&=\frac{1}{2}\tr\Big{\lbrace}D^{-1}\partial_{t}R\kk D^{-1}MD^{-1}M\Big{\rbrace}\ ,
\end{align}
where $R\kk^{(a)}=Z\tilde{R}_{z}^{(a)}$ and $R\kk^{(h)}=(M^{2}/4)\tilde{R}\kk^{(h)}$.
We need the expansion of every term in eq.~\eqref{N5} in second order in $A\mb$ or $F\mnb$. Since $M$ is linear in $F\mnb$ we can evaluate $\pi\kk^{(ah)}$ by setting $A\mb =0$ for the diagonal matrix $D$. The graphical representation of  $\pi\kk^{(ah)}$ is shown in Fig.~\ref{Fig.1}. We have
\noindent
indicated in Fig.\ref{Fig.1} the factors of $Z$, inferring $\pi_{k}^{(ah)}\sim Z$.

\begin{figure}[h]
\centering\includegraphics{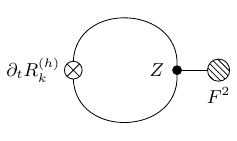}
\vspace*{-5pt}
\caption{Graviton fluctuation contributing to the flow of $Z$.}\label{Fig.2}
\end{figure}

For the contribution $\pi\kk^{(h)}$ we expand 
\bel{N6}
\pi\kk^{(h)}=\frac{1}{2}\tr\Big{\lbrace}P^{(h)-1}\partial_{t}R\kk^{(h)}\Big{\rbrace}+\pi\kk^{(h, F)}+\dots
\ee
with
\bel{N7}
\pi\kk^{(h,F)}=-\frac{1}{2}\tr\Big{\lbrace}P^{(h)-1}\partial_{t}R\kk^{(h)}P^{(h)-1}P^{(h,F)}\Big{\rbrace}\ .
\ee
The first term in eq.~\eqref{N6} yields the well known gravitational contribution to the flow of the scalar potential~\cite{PRWY, Wetterich:2019rsn}. The second term gives a contribution $\sim F\2$, as depicted in Fig.~\ref{Fig.2}. It is obvious that $\pi\kk^{(h,F)}$ is
\noindent
proportional to $Z$. Finally, $\pi\kk^{(a)}$ is the standard contribution from the fluctuations of gauge bosons, as computed in ref.~\cite{RW,CWG}. It is independent of $Z$. The measure contributions for the graviton fluctuations are independent of the background gauge field and only contribute to the flow of the scalar potential $V$ and $M\2$. The measure contributions from the gauge field fluctuations are the ones for Yang-Mills theories in flat space~\cite{RW,CWG}.

The simple one loop form of the flow equation which directly involves the physical fluctuations is another important advantage of the gauge invariant formulation of the flow equation. It renders the conceptual status of the flow equation for quantum gravity completely analogous to what is known for simpler quantum field theories, as scalar theories. We expect the result to be rather robust as long as the truncated propagators and vertices reproduce well the full propagators and vertices. The insertion of $\partial_tR_k$ renders the momentum integral in the loop ultraviolet finite, while the cutoff term $R_k$ in the inverse propagator removes possible infrared divergences. The momentum integrals are dominated by momenta $q^2\approx k^2$.

\section{Anomalous dimension for gauge\\couplings}\label{sec: ADGC}

From our discussion following eq.~\eqref{I19} it is clear that the sign and value of the gravity induced anomalous dimension $B_g$ is a key ingredient for understanding the scaling solution for gauge couplings. Asymptotic freedom is realized for $B_g>0$, while $B_g<0$ can lead to asymptotic safety. In the following we aim for a quantitative determination of $B_g$.

\subsection*{Expansion in small gauge couplings}

Summarizing the various contributions to the flow of $Z$ yields
\bel{N8}
\partial_{t}Z=-\eta_{F}Z=A_{g}Z+a_{4}\ ,
\ee
where the gauge boson contribution $a_{4}$ contains a part $\sim$~$\eta_{F}$ due to the $k$-derivative of the factor $Z$ in $R\kk^{(a)}$,
\bel{N9}
a_{4}=b_{F}+c_{F}\eta_{F}\ .
\ee
For SU($N$) Yang-Mills theories one has
\bel{N10}
b_{F}=\frac{11N}{24\pi\2}\; , \quad\; c_{F}=-\frac{21N}{96\pi\2}\ .
\ee
Similarly, the gravity induced anomalous dimension $A_{g}$ is given by 
\bel{N11}
A_{g}=b_{g}+c_{g}\eta_{F}\ ,
\ee
where $c_{g}$ arises from $\partial_{t}R\kk^{(a)}$ in the first diagram in Fig.~\ref{Fig.1}.
The coefficients $b_{g}$ and $c_{g}$ have to be computed (see below) from $\pi\kk^{(ah)}$ and $\pi\kk^{(h,F)}$ in eqs.~\eqref{N5} and~\eqref{N7}.

For $Z=1/g\2$ eq.~\eqref{N8} yields for $\eta_{F}$
\bel{N12}
\eta_{F}=-\frac{b_{g}+g\2 b_{F}}{1+c_{g}+g\2c_{F}}\ .
\ee
One infers the flow equation for the gauge coupling
\bel{N13}
\partial_{t}g\2 =\eta_{F}g\2=-\frac{b_{g}g\2 +b_{F}g^{4}}{1+c_{g}+c_{F}g\2}\ .
\ee
An expansion in small values of $g\2$ yields
\bel{N14}
\partial_{t}g\2=-B_{g}g\2-B_{F}g^{4}+\dots \ ,
\ee
with
\bel{N15}
B_{g}=\frac{b_{g}}{1+c_{g}}\;,\quad B_{F}=\frac{b_{F}}{1+c_{g}}-\frac{b_{g}c_{F}}{(1+c_{g})\2}\ .
\ee
In the absence of the metric fluctuations, $b_{g}=0$, $B_{g}=0$, $B_{F}=b_{F}$, the term $\sim g^{4}$ amounts to the standard one-loop $\beta$-function for gauge theories, while the next term $\sim g^{6}$ accounts for most of the two-loop contribution~\cite{RW,CWG}.
For $B_{g}>0$ the gauge coupling is asymptotically free. The gravitational contributions dominate for $g^{2}\rightarrow 0$.

\subsection*{Universality of gravity induced anomalous\\dimension}

There are two simple properties of the coefficients $b_{g}$ and $c_{g}$ that we can extract without detailed calculations. The first observes that $b_{g}$ is dimensionless. For $V=0$ the only scales in the truncation~\eqref{1} of the effective action are $M$ and $k$. In this limit $b_{g}$ is a function of $k\2/M\2$. The diagrams in Figs.~\ref{Fig.1},~\ref{Fig.2} involve the propagator $P^{(h)-1}$ of the metric fluctuations which is $\sim M^{-2}$. With $\partial_{t}R\kk^{(h)}\sim M\2$ one concludes
\bel{N16}
b_{g}=\tilde{b}_{g}\frac{k\2}{M\2}\;,\quad c_{g}=\tilde{c}_{g}\frac{k\2}{M\2}\ .
\ee
For $V=0$ both $\tilde b_g$ and $\tilde c_g$ are independent of $k$. This property is modified for $V\neq0$.
For $V\neq0$ the coefficients $\tilde{b}_{g}$ and $\tilde{c}_{g}$ will depend on the dimensionless combination
\bel{N17}
v=\frac{2V}{k^2M\2}.
\ee

Second, the coefficients $b_{g}$ and $c_{g}$ are universal in the sense that they do not depend on the particular gauge group. The gravity induced anomalous dimension is the same for all Yang-Mills theories, being equal to the one for an abelian model as quantum electrodynamics. For the contribution in Fig.~\ref{Fig.2} this follows directly from the observation that $P^{(hF)}$ only involves the gauge invariant bilinear of the field strength
\bel{N18}
\mathcal{F}^{\mu\rho\nu\sigma}=F_{z}{}^{\mu\rho}F^{z\nu\sigma}\ .
\ee
It is directly related to 
\bel{N19}
F_{z}\mno F\mnb^{z}=\eta\mnb\eta_{\rho\sigma}\mathcal{F}^{\mu\rho\nu\sigma}\ ,
\ee
without involving any particular relations for the generators of the gauge group.

For the first diagram in fig.~\ref{Fig.1} we write explicitly $\pi_k^{(ah)}=\pi_{k,1}^{(ah)}+\pi_{k,2}^{(ah)}$, where
\bel{N20}
\pi_{k,1}^{(ah)}=\frac{1}{2}\int_{q}\bigg{\lbrace}\Big{(}P^{(h)-1}\partial_{t}R\kk^{(h)}P^{(h)-1}\Big{)}_{\mu\nu\rho\sigma} C^{\rho\sigma\mu\nu}\bigg{\rbrace}
\ee
with
\bel{N21}
C^{\rho\sigma\mu\nu}=\Big{(} P^{(ha)}\big{(}P^{(a)}\big{)}^{-1}P^{(ha)\dagger}\Big{)}^{\rho\sigma\mu\nu} \ .
\ee
The momentum integral stands for $\int_{q}=\int d^{4} q/(2\pi)^{4}$. The gauge boson propagator is given by
\bel{N22}
\Big{(}\big{(}P^{(a)}\big{)}^{-1}\Big{)}_{\mn}^{zy}=Z^{-1}\big{(}q\2+\mathcal{R}_{k}(q)\big{)}^{-1}\tilde{P}\mnb \delta^{zy}\ ,
\ee
where $\mathcal{R}_{k}(q)$ is the cutoff function for the transversal gauge boson fluctuations. In analogy to the diagram in Fig.~\ref{Fig.2} one has
\begin{align}\label{N23}
C^{\rho\sigma\mu\nu}=&\frac{Z}{4}\big{(}q\2+\mathcal{R}_{k}(q)\big{)}^{-1}q_{\alpha}q_{\beta}\bigg{\lbrace}\Big{[}\mathcal{F}^{\alpha\rho\beta\mu}\tilde{P}^{\sigma\nu}\nn\\
&-\frac{1}{2}(\mathcal{F}^{\alpha\rho\beta\delta}\tilde{P}_{\delta}^{\sigma}\eta^{\mn}+\mathcal{F}^{\alpha\delta\beta\mu}\tilde{P}_{\delta}^{\nu}\eta^{\rho\sigma})\\
&+\frac{1}{4}\mathcal{F}^{\alpha\delta\beta\gamma}\tilde{P}_{\delta\gamma}\eta^{\rho\sigma}\eta^{\mn}+(\rho \leftrightarrow\sigma)\Big{]}+(\mu\leftrightarrow\nu)\bigg{\rbrace}\ ,\nn
\end{align}
involving only the gauge invariant combination~\eqref{N18}.

For the second diagram in Fig.~\ref{Fig.1},
\bel{N24}
\pi_{k,2}^{(ah)}=\frac{1}{2}\int_{q}\Big{\lbrace}\big{(}P^{(h)}\big{)}^{-1}_{\mn\rho\sigma}\tilde{C}^{\rho\sigma\mn}\Big{\rbrace}\ ,
\ee
one obtains $\tilde{C}^{\rho\sigma\mn}$ from $C^{\rho\sigma\mn}$
by replacing $(q\2+\mathcal{R}\kk)^{-1}$ by
\bel{47}
Z^{-1}\partial_{t}(Z\mathcal{R}\kk)(q\2+\mathcal{R}\kk)^{-2}=\gl\partial_t\mathcal{R}\kk-\eta_F\mathcal{R}\kk\gr\gl q\2+\mathcal{R}\kk\gr^{-2}\ .
\ee
The term $\sim\eta_{F}$ in eq.~\eqref{47} is responsible for the contribution in eq.~\eqref{N11}. We conclude that also the diagram Fig.~\ref{Fig.1} does not involve any particular relation for the generators of the gauge group. This establishes universality.

\subsection*{Computation of gravity induced anomalous\\dimension}

The quantitative values of the coefficients $b_{g}$, $c_{g}$ depend on the choice of the infrared cutoff function. We employ for the metric fluctuations the same cutoff function $\mathcal{R}\kk(q)$ as for the gauge bosons, with a structure adapted to the form of the inverse propagator
\bel{C1}
\big{(}R\kk^{(h)}\big{)}_{\mu\nu\rho\sigma}=\frac{M\2}{4}\mathcal{R}\kk(q)\big{(}\widehat{P}_{\mu\nu\rho\sigma}^{(t)}-2\widehat{P}_{\mu\nu\rho\sigma}^{(\sigma)}\big{)}\ .
\ee
The effect of the term $\mathcal{R}\kk^{(h)}$ in eq.~\eqref{leer02} is simply a replacement $q\2\rightarrow q\2+\mathcal{R}\kk(q)$ in the inverse metric propagator. By virtue of the projector properties the propagator for the physical metric fluctuations reads in presence of the infrared cutoff
\begin{align}\label{C2}
\big{(}P^{(h)}\big{)}_{\mu\nu\rho\sigma}^{-1}&=\frac{4}{M\2}\bigg{\lbrace}\big{(}q\2 +\mathcal{R}\kk (q)-vk\2\big{)}^{-1}\widehat{P}_{\mu\nu\rho\sigma}^{(t)} \nn\\
&-\frac{1}{2}\big{(}q\2 +\mathcal{R}\kk(q)-\frac{vk\2}{4}\big{)}^{-1}\widehat{P}_{\mu\nu\rho\sigma}^{(\sigma)}\bigg{\rbrace}\ ,
\end{align}
with $v$ given by eq.~\eqref{N17}.
Correspondingly, one obtains in the sector of physical metric fluctuations
\begin{align}\label{C4}
\big{(}P^{(h)-1}\partial_{t}R\kk ^{(h)}&P^{(h)-1}\big{)}_{\mn\rho\sigma}=\frac{4}{M\2}\big{(}\partial_{t}\mathcal{R}\kk (q)+(2-\eta_{g})\mathcal{R}\kk (q)\big{)}\nn\\
&\quad \times \Big{[}\big{(}q\2 +\mathcal{R}\kk (q)-vk\2\big{)}^{-2}
\widehat{P}_{\mn\rho\sigma}^{(t)}\\
&\quad-\frac{1}{2}\big{(}q\2 +\mathcal{R}\kk (q) -\frac{vk\2}{4}\big{)}^{-2}\widehat{P}_{\mn\rho\sigma}^{(\sigma)}\Big{]}\ .\nn
\end{align}
Here
\bel{C5}
\eta_{g}=-\partial_{t}\ln\big{(}\frac{M\2}{k\2}\big{)}\ .
\ee
reflects the possible $k$-dependence of $M\2$ in eq.~\eqref{C1}.

Taking things together we obtain explicit momentum integrals for the different contributions to the flow generator, as
\begin{align}\label{C6}
\pi\kk^{(ah)}=\frac{2}{M\2}\int_{q}\bigg{\lbrace}\Big{[}\mathcal{A}^{(t)}(q\2)+\mathcal{B}^{(t)}(q\2)\Big{]}C^{\rho\sigma\mn}\widehat{P}_{\mn\rho\sigma}^{(t)}\nn\\
-\frac{1}{2}\Big{[}\mathcal{A}^{(\sigma)}(q\2)+\mathcal{B}^{(\sigma)}(q\2)\Big{]}C^{\rho\sigma\mn}\widehat{P}_{\mn\rho\sigma}^{(\sigma)}\ ,
\end{align}
where
\bel{C7}
\mathcal{A}^{(t)}(q\2)=\big{(}\partial_{t}\mathcal{R}\kk +(2-\eta_{g})\mathcal{R}\kk\big{)}\big{(} q\2 +\mathcal{R}\kk -vk\2\big{)} ^{-2}\ ,
\ee
and
\bel{C8}
\mathcal{B}^{(t)}(q\2)=\big{(}\partial_{t}\mathcal{R}\kk -\eta_{F}\mathcal{R}\kk\big{)}\big{(} q\2 +\mathcal{R}\kk -vk\2\big{)} ^{-1}(q\2+\mathcal{R}\kk)^{-1}\ .
\ee
In the sector of physical scalar fluctuations of the metric $\mathcal{A}^{(\sigma)}$ and $\mathcal{B}^{(\sigma)}$ obtain from $\mathcal{A}^{(t)}$ and $\mathcal{B}^{(t)}$  by replacing $v\rightarrow v/4$. Similarly, one finds
\begin{align}\label{C9}
\pi\kk^{(hF)}=-\frac{2}{M\2}\int_{q}\bigg{\lbrace}&\mathcal{A}^{(t)} (q\2)P^{(hF)\mn\rho\sigma}\widehat{P}^{(t)}_{\rho\sigma\mu\nu}\\
& -\frac{1}{2}\mathcal{A}^{(\sigma)}(q\2)P^{(hF)\mn\rho\sigma}\widehat{P}_{\rho\sigma\mn}^{(\sigma)}\bigg{\rbrace}\ .\nn
\end{align}

We observe the opposite signs of the contributions $\pi\kk^{(ah)}$ and $\pi\kk^{(hF)}$. This important feature is at the origin of the possibility that the anomalous dimension for the gauge couplings can turn positive or negative, depending on the value of $v$. The opposite relative sign for the diagrams in Figs.~\ref{Fig.1} and~\ref{Fig.2} can also be seen by applying appropriate Feynman rules.

\subsection*{Gauge field configurations}

For the computation of $b_{g}$ and $c_{g}$ we choose a particular configuration of the gauge field 
\bel{G1}
A_{1}^{\overline{z}}=-\frac{F}{2}x_{2}\; , \quad A_{2}^{\overline{z}}=\frac{F}{2}x_{1}\ ,
\ee
for an arbitrary direction $\overline{z}$. The gauge field in all other color-directions vanishes, such that eq.~\eqref{G1} corresponds to an abelian gauge field. For constant $F$ the non-vanishing field strength components read 
\bel{G2}
F_{12}^{\overline{z}}=-F_{21}^{\overline{z}}=F\ ,
\ee
with
\bel{G3}
F^{z}{}_{\mn}F_{z}{}^{\mn}=2F\2\ .
\ee
Extracting the term $\sim F\2$ on the r.h.s. of the flow equation
\bel{G4}
\pi\kk=\frac{1}{2}\int_{x}s_{F}F\2+\dots \ ,
\ee
one obtains for the~\eqref{1}
\bel{G5}
\partial_{t}Z=s_{F}\ .
\ee

For the configuration ~\eqref{G1} one obtains
\bel{G6}
P^{(hF)\mn\rho\sigma}\widehat{P}_{\rho\sigma\mn}^{(t)}=\frac{ZF\2}{24}(20-10\Delta)\ ,
\ee
and
\bel{G7}
P^{(hF)\mn\rho\sigma}\widehat{P}_{\rho\sigma\mn}^{(\sigma)}=\frac{ZF\2}{24}(1-2\Delta)\ ,
\ee
where
\bel{G7A}
\Delta=\frac{q_{1}\2+q_{2}\2-q_{3}\2-q_{0}\2}{q\2}\ .
\ee

Similarly, for the contribution from Fig.~\ref{Fig.1} one needs
\bel{G8}
C^{\rho\sigma\mn}\widehat{P}_{\mn\rho\sigma}^{(t)}=\frac{5Z}{3\big{(}q\2+\mathcal{R}\kk(q)\big{)}}q_{\alpha}q_{\beta}\tilde{P}_{\rho\mu}\mathcal{F}^{\alpha\rho\beta\mu}\ ,
\ee
and
\bel{G9}
C^{\rho\sigma\mn}\widehat{P}_{\mn\rho\sigma}^{(\sigma)}=\frac{1}{20}C^{\rho\sigma\mn}\widehat{P}_{\mn\rho\sigma}^{(t)}\ .
\ee
For the configuration~\eqref{G1} we have
\bel{G10}
q_{\alpha}q_{\beta}\tilde{P}_{\rho\mu}\mathcal{F}^{\alpha\rho\beta\mu}=(q_{1}\2+q_{2}\2)F\2=\frac{1}{2}q\2 F\2(1+\Delta)\ .
\ee
Under the momentum integral we can employ
\bel{G11}
\int_{q}f(q\2)\Delta =0\ .
\ee

The independence of the expressions~\eqref{G6}-\eqref{G10} from the choice of gauge fields can be checked by assuming the existence of a different choice with
\bel{G11A}
\mathcal{F}^{\mu\rho\nu\sigma}=\frac{F\2}{6}(\eta^{\mn}\eta^{\rho \sigma}-\eta^{\mu\sigma}\eta^{\nu \rho})\ ,
\ee
where
\bel{G11B}
P^{(hF)\mn\rho\sigma}=\frac{ZF\2}{24}(2\eta^{\mu\rho}\eta^{\nu\sigma}+2\eta^{\mu\sigma}\eta^{\nu\rho}-\eta^{\mn}\eta^{\rho\sigma})\ ,
\ee
and
\begin{align}
\label{71A}
&\!\!C^{\mn\rho\sigma}=\frac{ZF\2q\2}{24(q\2+\mathcal{R}\kk)}\bigg{\lbrace}\Big{(}\big{[}2 \eta^{\rho\mu}\tilde{P}^{\sigma\nu}-\frac{1}{2}\tilde{P}^{\rho\sigma}\eta\mno -\frac{1}{2}\tilde{P}^{\mn}\eta^{\rho\sigma}\nn\\
&\;\; +\frac{3}{4}\eta\mno\eta^{\rho\sigma}-\tilde{P}^{\rho\mu}\tilde{P}^{\sigma\nu}\big{]}+(\rho\leftrightarrow\sigma)\Big{)}+(\mu\leftrightarrow\nu)\bigg{\rbrace} \ .
\end{align}
Thus the two gauge field configurations~\eqref{G2} and~\eqref{G11A} yield indeed the same formula for $s_F$ in eq.~\eqref{G4}.

\subsection*{Graviton domination}

Factoring out the factor $F^2$ we can now infer the coefficient $A_{g}$ in eq.~\eqref{N8},
\begin{align}\label{G13}
A_{g}=\frac{10}{3M\2}\int_{q}\bigg{\lbrace}&\frac{q\2}{q\2+\mathcal{R}\kk(q)}\Big{[}\mathcal{A}^{(t)}+\mathcal{B}^{(t)}\\
&-\frac{1}{40}(\mathcal{A}^{(\sigma)}+\mathcal{B}^{(\sigma)})\Big{]}-\mathcal{A}^{(t)}+\frac{1}{40}\mathcal{A}^{(\sigma)}\bigg{\rbrace}\ .\nn
\end{align}
One observes a marked dominance of the graviton (traceless transverse tensor) fluctuations. The subleading contribution from the scalar metric fluctuation proportional to 
$\mathcal{A}^{(\sigma)}$ and $\mathcal{B}^{(\sigma)}$ is suppressed for both diagrams by a factor around 1/40, unless $v$ takes large negative values. The graviton contributions are of a similar size for both diagrams, with a positive contribution $\sim \mathcal{A}^{(t)}+\mathcal{B}^{(t)}$ from Fig.~\ref{Fig.1}, and a negative contribution $\sim \mathcal{A}^{(t)}$ from Fig.~\ref{Fig.2}. Writing the graviton contribution in the form 
\bel{G14}
A_{g}^{(t)}=\frac{10}{3M\2}\int_{q}\bigg{\lbrace}\frac{q\2}{q\2+\mathcal{R}\kk(q)}\mathcal{B}^{(t)}-\frac{\mathcal{R}\kk(q)}{q\2+\mathcal{R}\kk(q)}\mathcal{A}^{(t)}\bigg{\rbrace}\ ,
\ee
we obtain for $v=0$, $\eta_{g}=0$, $\eta_{F}=0$
\bel{G15}
A_{g}^{(t)}=\frac{10}{3M\2}\int_{q}(q\2+\mathcal{R}\kk)^{-3}\Big{[}(q\2-\mathcal{R}\kk)\partial_{t}\mathcal{R}\kk-2\mathcal{R}\kk\2\Big{]}\ .
\ee
More generally, the sign of $A_{g}^{(t)}$ will depend on the value of $v$ and the precise shape of the cutoff function $\mathcal{R}\kk(q)$.

\subsection*{Numerical values of anomalous dimension}

For numerical values of the integrals $\int_{q}\mathcal{A}^{(t)}$ etc. we need to specify the form of the infrared cutoff function $\mathcal{R}\kk(q)$. We choose here the Litim cutoff~\cite{LIT}
\bel{N1.1}
\mathcal{R}\kk(q)=(k\2 -q\2)\theta(k\2-q\2)\ .
\ee
This restricts the momentum integral to $q\2<k\2$, and replaces in this range $q\2+\mathcal{R}\kk (q)$ by $k\2$.
The simple integration yields
\bel{N2.1}
\int_{q}\mathcal{A}^{(t)}=\int_{q}\frac{2k\2+(2-\eta_{g})(k\2-q\2)}{k^{4}(1-v)\2}=\frac{(8-\eta_{g})k\2}{96\pi\2(1-v)\2}\ ,
\ee
and
\bel{N3.1}
\int_{q}\mathcal{B}^{(t)}=\int_{q}\frac{2k\2-\eta_{F}(k\2-q\2)}{k^{4}(1-v)}=\frac{(6-\eta_{F})k\2}{96\pi\2(1-v)}\ ,
\ee
and similarly
\begin{align}\label{N4.1}
\int_{q}\frac{q\2}{q\2+\mathcal{R}\kk}\mathcal{A}^{(t)}&=\frac{(10-\eta_{g})k\2}{192\pi\2(1-v)\2}\; ,\nn\\ \int_{q}\frac{q\2}{q\2+\mathcal{R}\kk}\mathcal{B}^{(t)}&=\frac{(8-\eta_{F})k\2}{192\pi\2(1-v)}\ .
\end{align}

One obtains for $c_{g}$ in eq.~\eqref{N11}
\begin{equation}
c_{g}=-\frac{5k\2}{288\pi\2M\2}\Big{(}\frac{1}{1-v}-\frac{1}{40}\, \frac{1}{1-v/4}\Big{)}\ .
\end{equation}
This is typically a rather small quantity with only little influence on the running gauge couping in eq.~\eqref{N15}. In a good approximation we can identify $B_{g}=b_{g}$, $B_{F}=b_{F}$.
We extract our value for the gravity induced anomalous dimension
\begin{align}\label{N5.1}
b_{g}=&\frac{5k\2}{288\pi\2M\2}\bigg{[}\frac{8}{1-v}-\frac{6-\eta_{g}}{(1-v)\2}\nn\\
&-\frac{1}{5(1-v/4)}+\frac{6-\eta_{g}}{	40(1-v/4)\2}\bigg{]}\ .
\end{align}
For $\eta_{g}=0$ one finds vanishing $b_{g}$ for a critical $v_{c}$, which can be approximated by neglecting the scalar contribution, 
\bel{N6.1}
\eta_{g}(v_{c})=0\; ,\quad v_{c}\approx \frac{1}{4}\ .
\ee
For $v<v_{c}$ the gravity induced anomalous dimension is positive, $b_{g}>0$, while for $v>v_{c}$ one has $b_{g}<0$.

The dominant graviton part of our result~\eqref{N5.1} coincides with the result of ref.~\cite{CLPR}. It agrees with ref.~\cite{EV} once an error in the final formula of this paper is corrected. Larger differences are found as compared to the values of $b_{g}$ quoted in ref~\cite{DHR2}. This concerns both the overall size and the structure. Effects of different gauge fixing procedures are not expected to affect the contribution of the transverse traceless graviton fluctuations. They concern the scalar contribution which is found to be very small in our approach.
Using ``physical gauge fixing" the scalar contribution in ref.~\cite{CLPR} also coincides with our result. For more general gauge fixing the relative suppression of the scalar contribution as compared to the graviton contribution remains visible in ref.~\cite{CLPR}.
The dominant effect of different choices of infrared cutoff functions can be absorbed by a rescaling of the infrared cutoff scale $k$. Given that the results of refs.~\cite{CLPR},~\cite{EV} are obtained with different methods, the mutual agreement enhances confidence in the robustness of eq.~\eqref{N5.1}

\section{Standard model and grand\\unification}\label{sec: SMGU}

The dependence of the gravity induced anomalous dimension on the value of $v$ has important consequences for the ultraviolet behavior of the gauge couplings. The quantity $v$ involves the effective potential $V$ and the Planck mass $M$. Both are functions of the scalar field $\chi$ and we are interested here in the value for $\chi\to0$ or $\rhotil\to0$. The fixed point value of both $u=V/k^4$ and $w=M^2/(2k^2)$ for $\rhotil\to0$ depends on the particle content of transplanckian physics. Thus the question of asymptotic freedom or safety of the gauge couplings is influenced by this particle content. For the standard model one obtains asymptotic freedom of the gauge couplings. In contrast, for grand unified extensions one typically finds asymptotic safety provided that the number of scalar fields is not too large. For grand unification we will actually find an upper bound on the number of scalars. Beyond this bound the gauge couplings are predicted to vanish, in contrast to the non-zero values needed at the scale where gravity decouples.

\subsection*{Fixed points}

For a fixed point one has $M\2\sim k\2$ and $U\sim k^{4}$,
\begin{align}
\frac{M\2}{k\2}=2w_{*}\; &,\quad \eta_{ g}=0\ ,\nn\\
\!\frac{V}{k^{4}}=u_{*}\quad &, \quad v_{*}=\frac{u_{*}}{w_{*}}\ .
\end{align}
The gauge invariant flow equation with the same setting and infrared cutoff function yields in the graviton approximation~\cite{WY}
\begin{equation}
\label{Z0A}
w_{*}=\frac{1}{192\pi\2}\tilde{\mathcal{N}}_{M}+\frac{25}{128\pi\2(1-v_{*})}\ ,
\end{equation}
and~\cite{PRWY}
\begin{equation}
u_{*}=\frac{1}{128\pi\2}\tilde{\mathcal{N}}_{U}+\frac{5}{96\pi\2(1-v_{*})}\ ,
\end{equation}
where
\begin{align}\label{113A}
\tilde{\mathcal{N}}_{M}&=-N_{S}-N_{F}+4N_{V}+\frac{43}{6}\ ,\nn\\
\tilde{\mathcal{N}}_{U}&=N_{S}-2N_{F}+2N_{V}-\frac{8}{3}\ .
\end{align}
For massless particles $N_S$, $N_F$ and $N_V$ are the number of real scalars, Weyl fermions and gauge bosons. For nonzero particle masses these numbers are reduced by threshold functions.

The anomalous dimension at the fixed point depends on $w_*$ and $v_*$
\begin{equation}
b_{g}=\frac{5}{288\pi\2 w_{*}}\Big{(}\frac{4}{1-v_{*}}-\frac{3}{(1-v_{*})\2}\Big{)}\ .  
\end{equation}
Its sign only involves $v_*$. With
\begin{align}\label{Z1}
v_{*}=\frac{u_{*}}{w_{*}}&=\frac{3\tilde{\mathcal{N}}_{U}(1-v_{*})+20}{2\tilde{\mathcal{N}}_{M}(1-v_{*})+75}\nn\\
&=1-\frac{1}{4\tilde{\mathcal{N}}_{M}}\Big{\lbrace} 2\tilde{\mathcal{N}}_{M}-3\tilde{\mathcal{N}}_{U}-75\\
&\quad\quad +\sqrt{(2\tilde{\mathcal{N}}_{M}-3\tilde{\mathcal{N}}_{U}-75)^{2}+440\tilde{\mathcal{N}}_{M}}\Big{\rbrace}\ ,\nn
\end{align}
we obtain $b_{g}$ as a function of $N_{S}$, $N_{F}$ and $N_{V}$. For a given transplanckian particle content the gravity induced anomalous dimension is fixed.

As a first example we take the particle content of the standard model where~\cite{WY}
\begin{align}\label{Z2}
u_{*}&=-0.0507\quad ,\quad\quad w_{*}=0.00505\ ,\nn\\
v_{*}&=-10.05\ ,
\end{align}
and therefore
\bel{Z3}
b_{g}=0.118\ .
\ee
With a positive gravity induced anomalous dimension the gauge couplings are asymptotically free for the standard model coupled to gravity. The values of the gauge couplings cannot be predicted.

\begin{figure}[h]
\centering\includegraphics{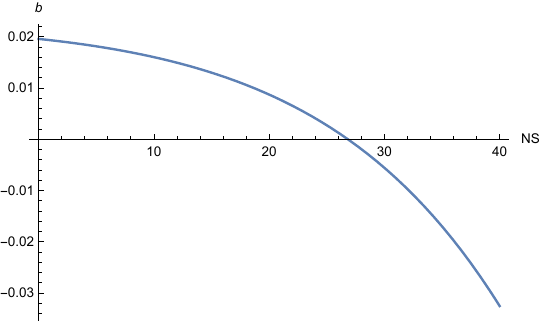}
\caption{Gravity induced anomalous dimension for $SO(10)$-unification. We plot the anomalous dimension $b_g$ as a function of the number of scalars $N_S$. It turns negative for $N_S>26$.}
\label{Fig.3}
\end{figure}

For an SO(10) grand unified theory (GUT) with $N_{F}=48$, $N_{V}=45$ we show the gravity induced anomalous dimension $b_{g}$ as a function of the number of scalars $N_{S}$ in Fig.~\ref{Fig.3}.
It turns negative for $N_{S}\gtrsim 30$ as a consequence of $v_{* }$ becoming larger than $1/4$. This is typically the case for a scalar sector leading to realistic spontaneous symmetry breaking. As long as $B_F$ remains positive the gauge coupling is asymptotically safe, with a fixed point value given by eq.~\eqref{I22}.

For $SO(10)$-unification one has
\bel{Z4}
B_{F}=\frac{50-\tilde{\mathcal{N}}_{10}}{16\pi\2}\ ,
\ee
and $\tilde{\mathcal{N}}_{10}$ given by the numbers $N_{R}$ of $R$-dimensional scalar-representations as~\cite{RS}
\begin{align}\label{Z5}
\tilde{\mathcal{N}}_{10}=\frac{1}{3}&\big{(}N_{10}+4N_{16}+8N_{45}+12N_{54}+28N_{120}+70N_{126}\nn\\
&+68N_{144}+56N_{210}\big{)}\ .
\end{align}
One finds an infrared unstable fixed point as long as $\tilde{\mathcal{N}}_{10}<50$. The gauge coupling is then asymptotically safe instead of asymptotically free. Again, its value cannot be predicted. In contrast, for a larger number of scalars $\tilde{\mathcal{N}}_{10}>50$ the gauge coupling becomes ´´trivial''. It is predicted to vanish for $\tilde{\rho}\approx 2w_{0}$, in contrast to observational requirements. Thus SO(10)-unification coupled to quantum gravity is viable only for $\tilde{\mathcal{N}}_{10}<50$.

As an example, we may consider a realistic $SO(10)$-model with scalars in a complex $10$ ($N_{10}=2$), complex $126$ ($N_{126}=1$) and real $54$ ($N_{54}=1$). With $\tilde{\mathcal{N}}_{10}=38$ this yields $B_F=3/(4\pi^2)$. The fixed point occurs for
\bel{Z5A}
g_*^2\approx\frac{5(v_*-\frac14)}{54w_*(1-v_*)^2}\ ,
\ee
with $v_*$ and $w_*$ given by eqs.~\eqref{Z1},~\eqref{Z0A} with $N_S=326$, $\tilde{\mathcal{N}}_{U}=320-8/3$, $\tilde{\mathcal{N}}_{M}=-194+43/6$.

\section{Conclusions}\label{sec: C}

We have performed a quantum gravity computation for the dependence of gauge couplings on a scalar field $\chi$. This prediction is based on the scaling solution for functional flow equations. No deviation from the scaling solution is necessary, such that fundamental scale invariance~\cite{CWFSI} can be realized. In the momentum region below the effective Planck mass the momentum dependence of the gauge couplings is compatible with the usual perturbative result for running couplings. 

In the limit of the infrared regulator scale $k$ going to zero the scaling solution predicts the quantum scale invariant standard model as an ``effective low energy theory" for momenta below the Planck mass. A small value of $k\approx 10^{-3}\text{eV}$, as advocated for interesting cosmologies with dynamical dark energy, induces only negligible modifications of the quantum scale invariant standard model, with a possible exception in the neutrino sector. The time variation of fundamental constants is negligibly small for present cosmology unless the neutrino sector yields an additional source. The same holds for apparent violations of the equivalence principle.
This renders dynamical dark energy with a very light scalar field compatible with observation without the need of any ``screening mechanism".

A more sizable time-variation of ``fundamental couplings" occurs in early cosmology when the ratio $\tilde{\rho}=\chi\2/k\2$ has been much smaller than today. This concerns the epoch of nucleosynthesis or before. We have estimated quantitatively the effect of the field dependence of the electromagnetic fine structure constant on the primordial element abundances generated by nucleosynthesis. The effect seems to be too small to be presently observable.

Finally, the scaling solution of quantum gravity imposes restrictions for model building. Coupled to a quantum field theory for the metric, grand unified theories based on the gauge groups SO(10) or SO(5) are viable only if the number of scalars is not too large. For these grand unified theories the gauge coupling is asymptotically safe, instead of asymptotic freedom realized for the standard model coupled to gravity.

In our truncation we have found no realization of the interesting case of an infrared stable fixed point at non-zero values of $g_*^2$. This fixed point would permit a prediction of the value of the gauge coupling near the Planck mass.

We believe that for the given truncation our result on the sign of the gravity induced anomalous dimension of the gauge coupling is rather robust. The question arises, however, if our truncation reflects well the dominant form of the graviton propagator in view of the substantial enhancement factor $(1-v_*)^{-2}$ for grand unified models. It seems well possible that higher derivative terms, as discussed in ref.~\cite{SWY}, play an important role for a large number of scalars. It seems not excluded that an infrared stable fixed point $g_*^2$ is found within an extended truncation.

Furthermore, there is no guarantee that the metric remains the basic degree of freedom for a quantum field theory or gravity. It seems likely to us that some form of gravity-induced anomalous dimension for the running gauge couplings applies to a more general class of theories. If the gravity induced anomalous dimension turns out to be positive for grand unified theories, due to an extended truncation or different degrees of freedom, a sufficiently large number of scalars in grand unified theories can turn $B_F$ to negative values. As a result the low energy value of the gauge couplings or the fine structure constant could be predicted.

%
%
%
%
%
\nocite{*} 
\bibliography{refs}
\end{document}